\documentclass[preprint,aps,nofootinbib]{revtex4}%
\usepackage{amsfonts}
\usepackage{amsmath}
\usepackage{amssymb}
\usepackage{graphicx}%
\begin{document}
\preprint{ }
\title{Minimal Dirac Fermionic Dark Matter with Nonzero Magnetic Dipole Moment}
\author{Jae Ho Heo}
\email{jheo1@uic.edu}
\affiliation{Physics Department, University of Illinois at Chicago, Chicago, Illinois
60607, USA }

\begin{abstract}
A neutral Dirac fermion $\psi$ is supplied as a singlet within the context of
the standard model (SM) and is considered as a dark matter (DM) candidate near
electroweak scale ($10\sim1000$ GeV) with nonzero magnetic dipole moment. The
Dirac particles have four different types of electromagnetic couplings (four
form factors) in general. We expect that the candidate mainly interacts with
SM particles through magnetic dipole moment (MDM), since MDM conserves the
discrete symmetries like parity (P), time reversal (T), and charge conjugation
(C) or its combination CP. The magnetic dipole moment constrained by the relic
density may be as large as 10$^{-18}\sim10^{-17}e$ $cm$. We show that the
elastic scattering is due to a spin-spin interaction for the direct detection,
and the predictions are under experimental exclusion limits of the current
direct detectors, XENON10 and CDMS II. We also consider the possibility of
WIMP detection in near future.

\end{abstract}

\maketitle

\section{Introduction}

The identity of the dark matter (DM) has been one of the most important
outstanding questions in physics and cosmology. It is known that DM must be
nonbaryonic, nonrelativistic (cold) and nonluminous matter \cite{Lange01}. So
it is not one of elementary particles contained in the standard moldel (SM).
The leading candidate from particle theory is a weakly interacting massive
particle (WIMP), since the annihilation cross section can lead to the correct
thermal residual abundance with the right magnitude. The new stable massive
particles could be generated in many extensions of the SM \cite{Gju96}%
\cite{Abi06}\cite{Mka05}\cite{Cpb01}\cite{Ygk07}.

In this letter we consider a Dirac fermionic dark matter candidate with mass
near electroweak scale (10$\sim$1000 GeV), which is supplied as a singlet
within the standard model context. The Dirac particles have four different
types of electromagnetic couplings (four form factors) in general. Among the
four couplings, we expect that magnetic dipole moment (MDM) is sizable for the
neutral fermion. The electric charges must be prohibited by the unbroken
$U(1)_{\text{Q}}$ gauge symmetry \cite{Max04} and the other couplings except
for magnetic dipole accompany violations of the discrete symmetries. The
violations of the discrete symmetries might be sizable for the DM, but the
violations have been estimated to be small for the known Dirac particles. We
assume that the candidate has the small violations of the discrete symmetries.

The singlet Dirac fermion $\psi$ should couple to the shifted photon
(hypercharge gauge boson $B$) near electroweak scale (10$\sim$1000 GeV),
instead of photon. The Lagrangian with the magnetic dipole moment of $B$,
$\mu_{B},$ is%

\begin{equation}
\mathcal{L}_{eff}=\frac{1}{2}\mu_{B}\overline{\psi}\sigma_{\mu\nu}\psi
B^{\mu\nu}%
\end{equation}

In the standard model context, the Lagrangian may be expressed about photon
and $Z$ boson.%

\begin{equation}
\mathcal{L}_{eff}=\frac{1}{2}\mu\overline{\psi}\sigma_{\mu\nu}\psi(F^{\mu\nu
}-\tan\theta_{W}Z^{\mu\nu})
\end{equation}
where $F^{\mu\nu}$ is the field strength for photon, $Z^{\mu\nu}$ for $Z$
boson and $\mu$ is the magnetic dipole moment\footnote{To be strict, this is a
magnetic dipole form factor, $F_{2}(q^{2})$. The magnetic dipole moment is
defined at $q^{2}=0$. However we do not distinguish the difference of both
terminologies in this letter.}.

\section{ANNIHILATION PROCESS}

In the leading order of the magnetic dipole moment, WIMP annihilates into SM
particles $(f\overline{f},WW,ZH)$ via the $s$-channel $\gamma,Z$ exchanges,
where the Feynman diagrams are given in Fig.1. The MDM operator is split into
energy and momentum dependent parts by the familiar Gordon identity.%

\begin{equation}
\overline{v}(p^{\prime})\sigma^{\mu\nu}q_{\nu}u(p)=i\overline{v}(p^{\prime
})(2M\gamma^{\mu}+(p^{\prime}-p)^{\mu})u(p)
\end{equation}

The MDM operator has energy dependence. This implies that there is no
threshold or helicity suppression due to violations of discrete symmetries.
There is also no lepton number violation for this annihilation process, since
the candidate is a Dirac particle. The below calculation for annihilation
rates confirm this fact.%

\begin{figure}
[ptb]
\begin{center}
\includegraphics[
trim=0.000000in 0.000000in 0.000000in -0.325624in,
height=1.0326in,
width=3.5129in
]%
{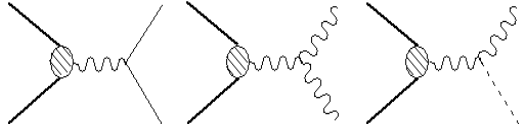}%
\caption{Feynman diagrams for WIMP pair annihilation into $f\overline
{f},WW,ZH$. The hatched circle indicates the vertex for the dipole coupling.}%
\label{FIG1}%
\end{center}
\end{figure}

The squared scattering amplitude in an average over spin of initial states and
a sum over colors and spins of final states is%

\begin{equation}
|\mathcal{M}|^{2}=\frac{N_{C}}{4}\mu^{2}\left\vert Q_{\gamma}P_{\gamma}%
-Q_{Z}\tan\theta_{W}P_{Z}\right\vert ^{2}L^{\mu\nu}Q^{\rho\sigma}g_{\mu\rho
}g_{\nu\sigma}%
\end{equation}
where $Q_{\gamma},Q_{Z}$ are couplings involved the outgoing particles with
photon and $Z$ boson, $P_{\gamma}\equiv1/s$, $P_{Z}\equiv1/\left[
(s-m_{Z}^{2})+im_{Z}\Gamma_{Z}\right]  $ imply photon and $Z$ boson
propagations, and $N_{C}$ is number of colors for the final states, 3 for
quarks and 1 for the other particles. $L^{\mu\nu}$ is the trace of initial
states and $Q^{\rho\sigma}$ stands for the trace of final states.

The consistant annihilation rates for each final state are
\begin{subequations}
\label{1}%
\begin{align}
\sigma_{f}v_{\text{rel}}  &  =%
{\displaystyle\sum\limits_{f}}
\frac{N_{C}\alpha\beta_{f}}{3}\left[  \left\vert e_{f}P_{\gamma}-\frac
{c_{V}P_{Z}}{2\cos^{2}\theta_{W}}\right\vert ^{2}\left(  s+2m_{f}^{2}\right)
+\left\vert \frac{c_{A}P_{Z}}{2\cos^{2}\theta_{W}}\right\vert ^{2}\left(
s-4m_{f}^{2}\right)  \right]  \cdot\mu^{2}(s+8M^{2})\\
\sigma_{W}v_{\text{rel}}  &  =\frac{\alpha\beta_{W}s}{24}\left\vert P_{\gamma
}-P_{Z}\right\vert ^{2}\frac{s^{2}+20sm_{W}^{2}+12m_{W}^{4}}{m_{W}^{4}}%
\cdot\mu^{2}(s+8M^{2})\\
\sigma_{ZH}v_{\text{rel}}  &  =\frac{\alpha\left\vert \boldsymbol{p}%
\right\vert }{2\sqrt{s}}\left\vert \frac{m_{Z}^{2}}{m_{W}\cos^{2}\theta_{W}%
}P_{Z}\right\vert ^{2}\left(  1+\frac{\boldsymbol{p}^{2}}{3m_{Z}^{2}}\right)
\cdot\mu^{2}(s+8M^{2})
\end{align}
where $v_{\text{rel}}$ is the relative velocity of the annihilating particles,
$\alpha$ is the electric fine structure constant, $\beta$ is the velocity of
the particle, $c_{V}=T_{f}^{3}-2e_{f}\sin^{2}\theta_{W},$ $c_{A}=T_{f}^{3}$
are the vector and axial couplings of $Z$-boson with electric charge $e_{f}$
and the third component of isospin $T_{f}^{3}$, \ and $\left\vert
\boldsymbol{p}\right\vert =\frac{1}{2\sqrt{s}}((s-m_{H}^{2}+m_{Z}^{2}%
)^{2}-4sm_{Z}^{2})^{1/2}$ is momentum of $Z$-boson or Higgs. The total
annihilation rate is $\sigma v_{\text{rel}}=\left(  \sigma_{f}+\sigma
_{W}+\sigma_{ZH}\right)  v_{\text{rel}}$. We show the explicit mass dependence
of fermions.%

\begin{figure}
[ptb]
\begin{center}
\includegraphics[
trim=0.000000in 0.000000in -0.221923in 0.000000in,
height=6.0231cm,
width=8.8326cm
]%
{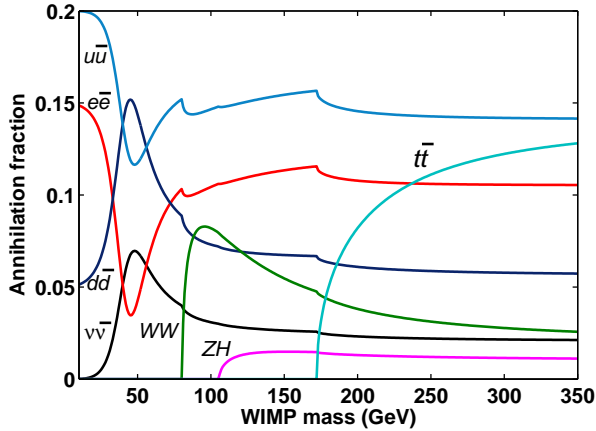}%
\caption{(color online) Annihilation fraction of particles. The fermionic
annihilation products dominate, hadronic annihilation products in particular.}%
\label{FIG2}%
\end{center}
\end{figure}

Fig.2 shows the annihilation fraction as a function of WIMP mass for the final
states. The fermionic annihilation products dominate, particularly hadronic
annihilation for entire parameter space. $ZH$ channel has the small
contribution since it interacts with WIMP via only $Z$-boson. The $\gamma-Z$
interference plays very important role for the $W$-boson channel\footnote{If
we only consider the interaction with photon, the contribution by $W$-boson
becomes extremely large in high energies. The interaction with $Z$ -boson
avoids this danger.}, since the $W$-boson channel contribution results in a
significant decrease in high energy due to the interference.

\section{Constraint from Relic Abundance}

In the very early universe, these particles would be present in large numbers
in thermal equilibrium. They could reduce their density only through pair
annihilation as universe is cool. This pair annihilation would be efficient
enough to reduce the present day number density of WIMP. The time evolution of
the relic density is described by the Boltzmann equation and we can solve the
equation in the approxmate way \cite{Ewk89}. After the freeze out, the actual
number of the dark matter per comoving volume\ becomes constant and the
present relic density is determined. The relic density is roughly%

\end{subequations}
\begin{equation}
\Omega_{\text{CDM}}h^{2}\simeq\frac{(1.07\times10^{9})x_{F}}{\sqrt{g_{\ast}%
}m_{Pl}(\text{GeV})\langle\sigma v_{\text{rel}}\rangle}%
\end{equation}
where $g_{\ast}$ is the effective degrees of freedom in equilibrium, $m_{Pl}$
is Planck mass, $x_{F}$ is the inverse freeze out temperature and
$\langle\sigma v_{\text{rel}}\rangle$ is the thermal average of annihilation rate.

The inverse freeze out temperature has almost no dependence on WIMP mass,
$x_{F}\simeq20$ near electroweak scale. The present relic density of CDM
measured from WMAP (Wilkinson microwave anisotropy probe) data \cite{dns07} is
$\Omega_{\text{CDM}}h^{2}=0.111\pm0.012$ at the level of 10\% accuracy. So the
thermal average of annihilation rate is $\langle\sigma v_{\text{rel}}%
\rangle\simeq0.62$ pb.

We take the thermal average for the annihilation obtained with
Maxwell-Botzmann statistics, which is described in Ref. \cite{pgo91}.
\begin{equation}
\langle\sigma v_{\text{rel}}\rangle=\frac{x}{8M^{5}K_{2}^{2}(x)}\int_{4M^{2}%
}^{\infty}ds\sigma(s)\left(  s-4M^{2}\right)  \sqrt{s}K_{1}\left(  \frac
{\sqrt{s}}{M}x\right)
\end{equation}
where $K_{i}$ is a modified Bessel function of order $i$.%

\begin{figure}
[ptb]
\begin{center}
\includegraphics[
trim=0.000000in 0.000000in -0.302740in 0.000000in,
height=5.9221cm,
width=8.8326cm
]%
{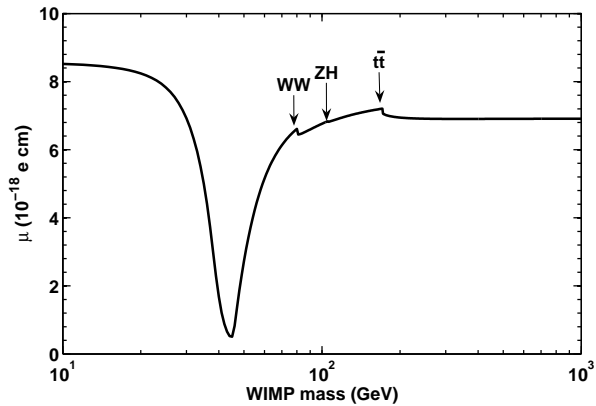}%
\caption{Allowed parameter sets from the relic abundance. The abundance is
chosen to be $\Omega_{\text{CDM}}h^{2}=0.111.$ The arrows indicate the points
where new channels start.}%
\label{FIG3}%
\end{center}
\end{figure}

Combination of Eq. (5) and (7) gives the allowed parameter sets for the
magnetic dipole moment of WIMP. Fig.3 shows the allowed parameter sets which
satisfy the WMAP measurements for relic density, in the case that the Dirac
fermions account for all the observed dark matter. So those become the upper
limits of WIMP magnetic dipole moment. All the annihilation rates of Eq.(5)
are included and the known inputs are used as well as Higgs mass of 120 GeV.
The anomaly indicates the resonant region of the $Z$ boson exchange where
$2M\simeq m_{Z}$. In this region, the dipoles are relatively small in order to
compensate the enhancement of annihilation rate by $Z$ boson resonace effect.
The WIMP magnetic dipole mmoment may be as large as $10^{-18}\sim10^{-17}e$
$cm$. If the dipole comes from new physics effect, we predict the new physics
with $O(\Lambda)\sim10$ TeV.

\section{DIRECT DETECTION}

The sensitivity of dark matter is controlled by their elastic scattering with
a nucleus in a detector. The Feynman diagram which give rise to the
WIMP-nucleon interaction is shown in Fig.4, which the $t$-channel exchanges of
$\gamma,Z$ are only possible. For the cold dark matter, the momentum transfer
is quite low, $|\boldsymbol{q}|$\ $\sim$ $O$(MeV)($E_{R}\sim$ $O$(KeV)). The
interaction via $Z$ boson is $1/m_{Z}^{2}\sim G_{F}$ where $G_{F}$ is the
Fermi constant, and $1/\boldsymbol{q}^{2}$ for the photon. The interaction via
$Z$ boson may be neglected. We thus only consider the interaction via photon.%

\begin{figure}
[ptb]
\begin{center}
\includegraphics[
trim=0.000000in 0.000000in -0.021613in 0.000000in,
height=3.5541cm,
width=8.0792cm
]%
{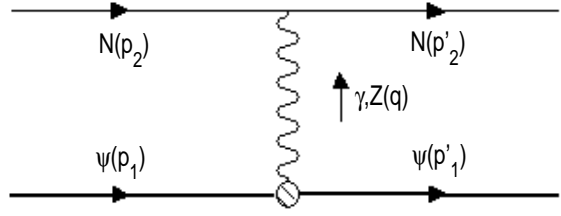}%
\caption{Feynman diagram relevant to WIMP-nucleon elastic scattering. The
hatched circle indicates the vertex for the dipole coupling.}%
\label{FIG4}%
\end{center}
\end{figure}

\subsection{Elastic-scattering cross section}

The effective Lagrangian may be described by four fermion interaction,%

\begin{equation}
\mathcal{L}_{q}=i\frac{\mu ee_{q}}{\boldsymbol{q}^{2}}\left(  \overline{\psi
}\sigma^{\mu\nu}q_{\nu}\psi\right)  \left(  \overline{q}\gamma_{\mu}q\right)
\end{equation}
where $e$ is the electric coupling and $e_{q}$ is an electric charge for the
quark $q$.

The MDM operator $\overline{\psi}\sigma^{\mu\nu}q_{\nu}\psi$ has no energy
dependence for the elastic scattering. So we take only space component of the
operator and the leading order of $|\boldsymbol{q}|$ in the nonrelativistic
expansion of the spinor. It results in the spin-spin insteraction between WIMP
and quark.

The effective Lagrangian for nucleons may be expressed by the spin distrbution
of quarks.%

\begin{equation}
\mathcal{L}_{N}\supset2iM\frac{\mu ea_{N}}{\boldsymbol{q}^{2}}\left(
\overline{\psi}\gamma^{i}\psi\right)  \left(  \overline{N}\gamma_{i}N\right)
\end{equation}
where the symbol $N$ stands for proton ($p$) or neutron ($n$), and
$a_{N}\equiv%
{\textstyle\sum}
e_{q}\Delta^{(N)}q$ where $\Delta^{(N)}q$ is the fraction of nucleon spin
carried by quarks of type $q$. The fractions for proton, according to
Ref.\cite{jre00}, are $\Delta^{(p)}u=0.78,\Delta^{(p)}d=-0.48$, and
$\Delta^{(p)}s=-0.15$, each with an uncertainty of about $\pm0.02$, and
$\Delta^{(n)}u=\Delta^{(p)}d,\Delta^{(n)}d=\Delta^{(p)}u$, $\Delta
^{(n)}s=\Delta^{(p)}s$ for neutron.

The differential cross section of WIMP-nucleus in a spin average of initial
states and a spin sum of final states is given by%

\begin{equation}
\frac{d\sigma_{el}}{d\boldsymbol{q}^{2}}=\frac{\mu^{2}a_{N}^{2}}{4\pi
v^{2}\boldsymbol{q}^{4}}\left(  \frac{e}{2m}\right)  ^{2}\frac{1}%
{(2S+1)(2J+1)}%
{\textstyle\sum\limits_{j,j^{\prime},s,s^{\prime}}}
|\mathcal{M}|^{2}%
\end{equation}
where $S=\frac{1}{2}$ is the WIMP spin, $J$ is the nuclear spin and $m$ is
mass of the atomic nucleus.

The squared scattering amplitude can split into two parts, WIMP and nuclear
parts. The WIMP part is%

\[
\frac{1}{(2S+1)}%
{\textstyle\sum\limits_{s,s^{\prime}}}
|\mathcal{M}|_{\text{WIMP}}^{2,ij}=4\frac{S(S+1)}{3}\left(  \boldsymbol{q}%
^{2}\delta^{ij}-q^{i}q^{j}\right)
\]

The squared amplitude for the nuclear part depends on the nuclear structure
function, $S(|\boldsymbol{q}|)$, which is normalized by the nuclear spin.%

\begin{equation}
\frac{1}{(2J+1)}%
{\textstyle\sum\limits_{j,j^{\prime}}}
|\mathcal{M}|_{\text{nucleus}}^{2,ij}=\frac{J(J+1)}{3}S(|\boldsymbol{q}%
|)\left(  \boldsymbol{q}^{2}\delta^{ij}-q^{i}q^{j}\right)
\end{equation}

The nuclear structure function $S(|\boldsymbol{q}|)$ may be expressed in terms
of three form factors, isoscalar ($a_{0}$), isovector ($a_{1}$) and the
inteference ($a_{0}a_{1}$).%

\begin{equation}
S\left(  |\boldsymbol{q}|\right)  =a_{0}^{2}S_{00}\left(  |\boldsymbol{q}%
|\right)  +a_{0}a_{1}S_{01}\left(  |\boldsymbol{q}|\right)  +a_{1}^{2}%
S_{11}\left(  |\boldsymbol{q}|\right)
\end{equation}
with $a_{0}=a_{p}+a_{n}$ and $a_{1}=a_{p}-a_{n}$. The functions $S_{ij}$
encompass the magnitude of the spin associated with the nucleon populations,
as well as the effects of the spatial distribution of that spin at nonzero
momentum transfers.

The WIMP-nucleus differential cross section results in%

\begin{equation}
\frac{d\sigma_{el}}{d\boldsymbol{q}^{2}}\ =\frac{\mu^{2}\mu_{A}^{2}}{6\pi
v^{2}}\frac{J+1}{J}F^{2}(|\boldsymbol{q}|)
\end{equation}
where $\mu_{A}^{2}\equiv\left(  \frac{e}{2m}J\right)  ^{2}S(0)$ is the
magnetic dipole moment of the atomic nucleus and $F^{2}(|\boldsymbol{q}%
|)\equiv S(|\boldsymbol{q}^{2}|)/S(0)$ is the normalized form factor,
$F^{2}(0)=1$. We have to integrate over $\boldsymbol{q}^{2}$ for the actual
cross section. There are two quantities which depend on atomic nuclei, nuclear
spin $J$ and the nuclear structure function $S(|\boldsymbol{q}^{2}|)\sim
\mu_{A}^{2}F^{2}(|\boldsymbol{q}|)$. Those become the reasons we have the
different signatures for the various targets.

The cross section at $|\boldsymbol{q}|=0$ is%

\begin{equation}
\sigma=%
{\displaystyle\int\nolimits_{0}^{4M_{r}^{2}v^{2}}}
\frac{d\sigma_{el}(|\boldsymbol{q}|=0)}{d\boldsymbol{q}^{2}}d\boldsymbol{q}%
^{2}=\frac{2}{3\pi}M_{r}^{2}\mu^{2}\mu_{A}^{2}\frac{J+1}{J}%
\end{equation}
where $M_{r}=Mm/\left(  M+m\right)  $ is the reduced WIMP-nucleus mass.

The WIMP-nucleon cross sections result in%

\begin{equation}
\sigma_{p,n}=\frac{1}{2\pi}m_{r}^{2}\mu^{2}\left(  \frac{ea_{p,n}}{2m}\right)
^{2}\frac{J+1}{J}%
\end{equation}
where $m_{r}=Mm_{N}/(M+m_{N})$ is the reduced WIMP-nucleon mass. This cross
section still includes one of the target properties, nuclear spin $J$, so it
is not totally the nucleus independent quantity.

\subsection{WIMP detection}

The expected event rate depends on WIMP-nucleus cross section and the WIMP
flux on the Earth, so the differential event rate per unit target mass and
unit time is%

\begin{equation}
\frac{dR}{d\boldsymbol{q}^{2}}=\frac{\rho_{D}}{mM}\int vf(v)\frac{d\sigma
_{el}}{d\boldsymbol{q}^{2}}dv
\end{equation}
where $\rho_{D}\simeq0.3$ GeV/cm$^{3}$ is the local DM density in the solar
vicinity and $f(v)$ is the distribution of speeds relative to the dectector.
We take the Maxwellian distribution considered the annual modulation
\cite{kath88}.%

\begin{equation}
f(v)dv=\frac{v}{v_{E}v_{0}\sqrt{\pi}}\left\{  \exp\left[  -\frac{\left(
v-v_{E}\right)  ^{2}}{v_{0}^{2}}\right]  -\exp\left[  -\frac{\left(
v+v_{E}\right)  ^{2}}{v_{0}^{2}}\right]  \right\}  dv
\end{equation}
where $v_{0}=220$ km/s is the circular speed of the Sun around the Galactic
center and $v_{E}\sim v_{0}$ is the Earth speed to the Sun. $v_{E}$ changes as
the Earth's motion come into and out of alignment with the Sun's motion around
the Galaxy. This causes an yearly modulation in the event rate which peaks
around June 2nd each year \cite{kgr88}. We take the average velocity,
$v_{E}=232$ km/s.

With the relation $\boldsymbol{q}^{2}$ $=2mE_{R}$, Eq. (13), (16) and (17)
lead to the differential event rate.%

\begin{equation}
\frac{dR}{dE_{R}}=\frac{\rho_{D}\mu^{2}\mu_{A}^{2}}{3\pi M}\frac{(J+1)}%
{J}F^{2}(E_{R})\int_{v_{\min}}^{\infty}\frac{f(v)}{v}dv\text{\ }%
\end{equation}

The final formula for the event rate per unit target mass and unit time is
given by%

\begin{equation}
R=\frac{\rho_{D}\sigma}{2MM_{r}^{2}}\int_{E_{R,\min}}^{E_{R,\max}}dE_{R}%
F^{2}(E_{R})\cdot\frac{1}{2v_{E}}\left[  \operatorname{erf}\left(
\frac{v_{\min}+v_{E}}{v_{0}}\right)  -\operatorname{erf}\left(  \frac{v_{\min
}-v_{E}}{v_{0}}\right)  \right]
\end{equation}
where $v_{\min}=\sqrt{\frac{E_{R,\min}m}{2M_{r}^{2}}}$. The form factor has to
be integrated over $E_{R}$ in the recoil energy range. The form factors for
the spin-spin interaction are relatively slowly decreased, so it results in
suppressions for the event rates.

\begin{table}[t]
\caption{Current and planned Dark Matter detectors. XENON10 and CDMS II are
the most sensitive detectors and SuperCDMS is a future detector.}%
\label{table1}
\begin{ruledtabular}
\begin{tabular}{ccccc}
$\text{Experiment}$ & $\text{Recoil energy range}$  & $\text{Target}$& $\text{Spin}$ & $\text{
Mass(Kg)}$ \\
\hline
$\text{XENON10}$ & $4.5\sim 27\text{KeV}$ & $^{131}\text{Xe}$&${3 \over 2}$ & $5.4\text{Kg}$ \\
$\text{CDMS II}$ & $10\sim 100\text{KeV}$ & $^{73}\text{Ge}$&${9 \over 2}$& $100\text{Kg}$
\\
$\text{SuperCDMS}$ & $15\sim 45\text{KeV}$ & $^{73}\text{Ge}$&${9 \over 2}$ & $100\text{
Kg}$\\
\end{tabular}
\end{ruledtabular}\end{table}

There are various isotope targets which have been considered in WIMP
detection. Among them, we consider the isotope targets, $^{73}$Ge and $^{131}%
$Xe, which have been used at the most sensitive detectors, CDMS II and
XENON10. The recent estimated experimental sensitivities are 0.6 event/kg/day
for XENON10 \cite{xenon08} and CDMS II \cite{cdm08}, and the planned
experimental sensitivity is $1\times10^{-4}$ event/kg/day at SuperCDMS
\cite{scdm06}. The recoil energy ranges are listed at Table II for the
corresponding detectors.

Fig.5 shows the WIMP-nucleon cross sections from Eq.(15) with the experimental
limits in the WIMP mass range $10\sim1000$ GeV. The predicted curves and the
experimental limit contours are for pure proton coupling $(a_{n}=0)$ and pure
neutron coupling $(a_{p}=0)$. The predicted curves show a little different
from both isotope targets because of the nuclear spin effect. The form factors
related to the spin structure functions are followed by Dimitrov \textit{et
al.}\cite{vim95} for $^{73}$Ge nucleus and Ressell \textit{et al.}\cite{mtr97}
for $^{131}$Xe nucleus, which are consistent with the experimental
measurements for MDM in the reasonable precision. The $^{73}$Ge and $^{131}$Xe
nuclei both contain an unpaired neutron, so the sensitivity to pure-neutron
coupling is stronger. The predictions are under the current discovery limits
for the entire mass range and are relatively close to the experimental limits.
We should not discard the WIMP discovery through the magnetic dipole moment in
near future.%

\begin{figure}
[ptb]
\begin{center}
\includegraphics[
trim=0.000000in 0.000000in -0.361400in 0.000000in,
height=8.8831cm,
width=13.25cm
]%
{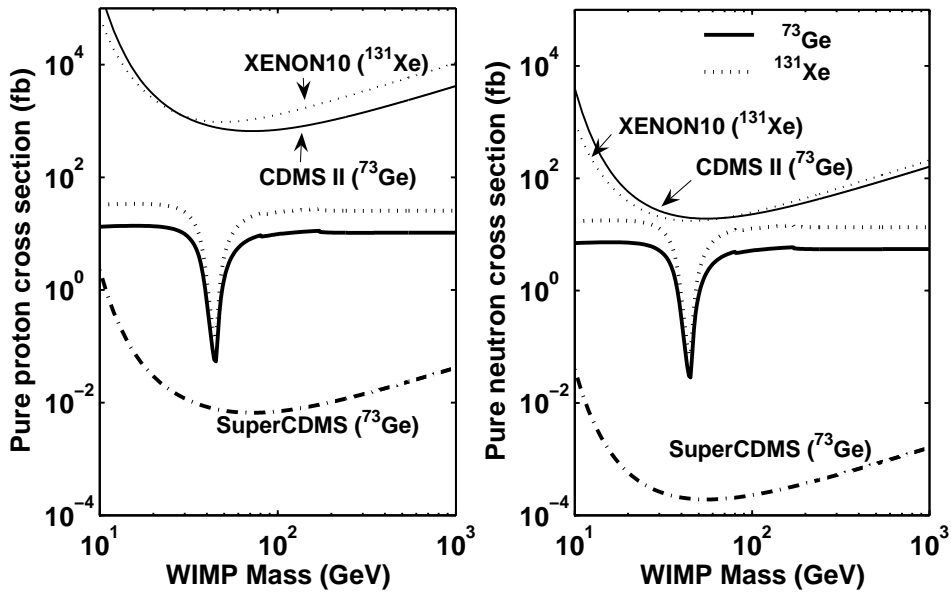}%
\caption{The predictions for WIMP-nucleon cross sections, $\sigma_{p,n}$, as a
function of WIMP mass. The light dotted lines are the exclusion limits from
the XENON10 experiment and the solid lines from CDMS II experiment. Also shown
by dash-dotted lines are the experimental sensitivities of the planned
detector SuperCDMS, respectively.}%
\label{FIG5}%
\end{center}
\end{figure}

\section{Conclusion and Discussion}

A Dirac fermion which was supplied as a singlet within the standard model
context may mainly interact with the ordinary matters through magnetic dipole
moment (MDM). It must be more general that the candidate couples to the
shifted photon (hypercharge gauge boson) than coupling to photon directly near
electroweak scale $10\sim1000$ GeV. Assuming that the Dirac fermions account
for all the observed dark matter, the MDM constrained by the relic density may
be as large as $10^{-18}\sim10^{-17}e$ $cm$. The WIMP with nonzero MDM
interacts with targets through MDM of the target nuclei and the cross sections
are under the current experimental limits. Since the preditions are close to
the current experimental limits, the detection of the WIMP through MDM should
be considered be in near future.

Besides the direct search of WIMP, there are several experiments to search for
dark matter indirectly, such like neutrino telescopes, $\gamma$-ray telescopes
and antimatter telescopes. They might give the more serious constraints
\cite{heo09}, but the indirect searches for dark matter accompany uncertain
informations in astrophysics. This scenario, which the dark matter has Dirac
fermionic nature and mainly interacts with the ordinary matters through the
magnetic dipole moment, might not be ruled out.

We have also studied the possibility of WIMP detection with the method of a
photon or $jet$ tagging in ILC and LHC. Several tens or hundreds of WIMP
events are expected, but these signals suffer the huge irreducible background
come from the neutrino channel. So the expected signals are swamped by the
irreducible background.


\begin{thebibliography}{99}                                                                                               %


\bibitem {Lange01}A. Lange \textit{et al.}, Phys. Rev. D \textbf{63}, 042001
(2001) [arXiv:astro-ph/0005004 ] ; N.W. Halverson \textit{et al}., Astrophys.
J. Suppl. \textbf{148}, 175 (2003) [arXiv:astro-ph/0302209]

\bibitem {Gju96}G. Jungman, M. Kamionkowski and K. Griest, Phys. Rep.
\textbf{267,} 195 (1996) ; K. Griest and M. Kamionkowski, Phys. Rept.
\textbf{333}, 167 (2000)

\bibitem {Abi06}A. Birkedal, A. Noble, M. Perelstein and A. Spray, Phys. Rev.
D \textbf{74}, 035002 (2006) [arXiv:hep-ph/0603077]

\bibitem {Mka05}M. Kakizaki, S. Matsumoto, Y. Sato and M. Senami, Phys. Rev. D
\textbf{71}, 123522 (2005) [arXiv:hep-ph/0508283]

\bibitem {Cpb01}C.P. Burgess, M. Pospelov and T. Veldhuis, Nucl. Phys.
\textbf{B619}, 709 (2001) [arXiv:hep-ph/0011335]; J. McDonald, Phys. Rev. D
\textbf{50}, 3637-3649 (1994) [arXiv:hep-ph/0702143]

\bibitem {Ygk07}Y.G. Kim, and K.Y. Lee, Phys. Rev. D\textbf{ 75,} 115012
(2007) [arXiv:hep-ph/0611069] ; Y.G. Kim, K.Y. Lee and S. Shin, J. High Energy
Phys. 0805 100 (2008) [arXiv:hep-ph/0803.2932]

\bibitem {Max04}M. Dvornikov and A. Studenikin, Phys. Rev. D\textbf{ 69},
073001 (2004); L.G. Cabral-Rosettic, J. Bernabeu, J. Vidal and A. Zepeda, Eur.
Phys. J. C\textbf{ 12}, 633 (2000); A. Denner, G. Weiglein and S. Dittmaier,
Nucl. Phys. \textbf{B440}, 95 (1995)

\bibitem {Ewk89}E.W. Kolb and M.S. Turner, The Early Universe (Addison-Wesley
Redwood City, 1989)

\bibitem {dns07}D.N. Spergel \textit{et al}., (WMAP Collaboration), Astrophys.
J Suppl. \textbf{170}, 377 (2007) [arXiv:astro-ph/0603449]

\bibitem {pgo91}P. Gondolo and G. Gelmini, Nucl. Phys. \textbf{B360, }145 (1991)

\bibitem {jre00}J.R. Ellis, A. Ferstl and K.A. Olive, Phys. Lett. B
\textbf{481}, 304 (2000) [arXiv:hep-ph/0001005]

\bibitem {kath88}K. Freese, J. Frieman and A. Gould, Phys. Rev. D \textbf{37},
3388 (1988)

\bibitem {kgr88}K. Griest, Phys. Rev. D\textbf{ 37}, 2703 (1988); K. Freese,
J.A. Frieman, and A. Gould, Phys. Rev. D \textbf{37}, 3388 (1988) ; A.
Drukier, K. Freese and D. Spergel, Phys. Rev. D\textbf{ 33}, 3495 (1986)

\bibitem {xenon08}J. Angle \textit{et al. }(XENON10 Collaboration)\textit{,}
Phys. Rev. Lett. \textbf{101}, 091301 (2008) [arXiv:astro-ph/0802.3530]

\bibitem {cdm08}Z. Ahmed \textit{et al. }(CDMS Collaboration)\textit{,} Phys.
Rev. Lett. \textbf{102}, 011301 (2009) [arXiv:astro-ph/0802.3530] ; Phys. Rev.
D \textbf{73}, 011102 (2006) [arXiv:astro-ph/0509269]

\bibitem {scdm06}D.S. Akerib \textit{et al. }(CDMS Collaboration)\textit{,}
Nucl. Instrum. Meth. A \textbf{55}, 411 (2006)\textit{ }[arXiv:astro-ph/0503583]

\bibitem {vim95}V.I. Dimitrov, J. Engel and S. Pittel, Phys. Rev. D
\textbf{51}, R291 (1995) ; J. Engel, Phys. Lett. \textbf{B264}, 114 (1991) ;
M.T. Ressel \textit{et al., }Phys. Rev. D \textbf{48}, 5519 (1993)

\bibitem {mtr97}M.T. Ressell and D.J. Dean, Phy. Rev. C \textbf{56}, 535
(1997) [arXiv:hep-ph/970229] ; M.A. Nikolaev and H.V. Klapdor-Kleingrothaus,
Z. Phys. A \textbf{345}, 183 (1993)

\bibitem {heo09}proceeding
\end{thebibliography}
\end{document}